\documentstyle[aps,preprint]{revtex}
\draft	
\begin{document}
\title{Gauge Invariant Bosonization of \\
Quantum Hall Systems and Skyrmions : Kinematics}
\author{ Sreedhar B. Dutta and R. Shankar}

\address{Institute of Mathematical Sciences, CIT Campus, 
Chennai (Madras) 600113, }

\date{\today}

\maketitle
\begin{abstract}
We develop a systematic semiclassical approximation scheme for quantum
Hall skyrmions near filling factors $\nu = {1 \over 2n+1}$, which is 
exact in the long wavelength limit. We construct a coherent state basis for 
the Hilbert space of Chern-Simons gauge fields and composite bosons with spin. 
These states are projected to the physical gauge invariant subspace and their 
wavefunctions explicitly evaluated. The lowest Landau level (LLL) condition is 
shown to be equivalent to an analyticity condition on the parameters. 

The matrix elements of physical observables between these states are shown to be 
calculable in the limit of small amplitude long wavelength density fluctuations. 
The electric charge density is shown to be proportional to the toplological 
charge density if and only if the LLL condition is satisfied. 

We then show that these states themselves form a generalised coherent state
basis, parameterised by the values of physical observables. The theory can 
therefore be written in terms of these gauge invariant bosonic fields in the 
long wavelength regime. The off diagonal matrix elements of observables in 
these coherent states are computed and shown to vanish in the long
wavelength limit. Thus we are able to prove that the classical description
of the skyrmion is exact in the limit of large skyrmions.

\end{abstract}
\newpage
\section{Introduction}
\label{intro}

The combination of a low value of effective mass and a low value of the Zeeman
coupling in GaAs makes the spin degree of freedom relevent for quantum Hall 
systems in this material. Initial calculations by Chakraborty and Zhang 
\cite{tapash} showed that the low energy quasiparticles in such systems were
spin reversed. Kane and Lee \cite{klee} then argued that the quasiparticles
could be extended spin textures that are described by topological solitons
of the non-linear sigma model (NLSM): skyrmions. Sondhi et. al. \cite{sondhi} 
estimated the energy of skyrmions using an effective NLSM and showed that they 
are energetically favoured over single spin reversed quasiparticles. This was 
corroborated by Hartree-Fock calculations by Fertig et. al. \cite{fertig}.

Experimental evidence for
the quasiparticles being extended spin textures initially came from Knight 
shift measurements of Barret et. al.\cite{barret}, optical magneto-absorption 
experiments of Aifer et. al. \cite{goldberg} and tilted field transport 
measurements of Schrieffer et. al \cite{schrieffer}. These experiments found 
the average value of the spin of the quasiparticle to be $\sim 2-7$ depending 
on the system parameters. Subsequently, experiments have been done in systems 
where the Land\'e $g$ factor is reduced by high pressures \cite{press} or by 
engineering the material \cite{shaygan} yielding evidence of quasiparticles
with very large, ($\sim 20 - 30$), values of spin. A negative result has also
been reported by Kukushkin et. al.\cite{kukushkin}, where optical 
measurements found the quasiparticles to have spin 
${1 \over 2}$, i.e. no spin textures. There has been no explanation of this 
result in the literature so far.

In the NLSM approach, the quasiholes (particles) are assumed to be described by the 
classical solutions of the model. The electrical charge density is assumed to
be equal to the topological charge density, thus the quasiparticles are the
classical solutions in in the toplological charge, $Q_{top} = 1$, sector.
Near $\nu = 1$, the NLSM energy functional and the relation between the 
topological charge density and electrical charge density has been derived by 
many workers in the LLL, long wavelength approximation 
\cite{sondhi,bychkov,rray}. 

What is the regime of validity of this model and is there a limit where the 
classical approximation is exact ? This is the main question we address and 
answer in this paper. This question has previously been discussed by Girvin 
et. al. \cite{girvin}, where they numerically compare the skyrmion energies
form the NLSM, Hartree-Fock and exact diagonalisation. The energies match in
the limit of large sized skyrmions (with size $R \ge 15 l_c,~l_c$ being the
magnetic length. This calculation suggests that the classical approximation
may be exact in long wavelength limit. In this paper we start from the
microscopic theory and analytically address the following questions:

\noindent
1(a). When is the topological charge proportional to the electric charge ?\\ 
1(b). When are the  corresponding densities proportional to each other ?

\noindent
2. What is the limit in which the classical approximation is exact ?

These questions automatically throw up a third one:

\noindent
3. How can the LLL condition be imposed in the classical theory ?

The composite boson formalism \cite{kivelson} is the obvious choice of
the formalism to use to address the question of the classical limit.
However, we would also like to impose the LLL condition, which is a condition 
on the quantum states. We therefore develop a coherent state formalism
for composite bosons.

In section \ref{cbtcs}, we construct a coherent state basis for the
Hilbert space of the bosons and the Chern-Simons gauge fields. The
anticommuting electron operators are then constructed in terms of the
bosonic fields. This construction provides the explicit mapping between
the electronic Hilbert space and operators and the gauge invariant
sector of the composite boson theory. 

Section \ref{gibs} concentrates on
the gauge invariant sector of the theory. The coherent states are
projected on to the gauge invariant sector and their wave functions
calculated. We then find that the wave functions thus obtained are
exactly of the form written down previously by Ezawa \cite{ezawa}. The LLL
condition is easily seen to be equivalent to an analyticity condition on
the parameters labelling the coherent states. The coherent states are
labelled by the values of the bosonic spinor field $\phi_\sigma(x)$
and the Chern-Simons gauge field $\alpha_i(x)$. The states projected to
the gauge invariant sector depend only on gauge invariant combinations
of these fields i.e the longitudinal part of $\alpha(x)$ gets related 
to the phase of $\phi(x)$. An inspection of the wavefunctions reveals another
local invariance that relates the transverse part of $\alpha(x)$ to the
magnitude, $\phi^\dagger(x)\phi(x)$ of the bosonic fields. This transformation 
is not unitarily realised. Nevertheless it implies that
the physical states can be labelled by a single bosonic spinor field
which we denote by $W_\sigma(x)$. We then show that the matrix elements
of the observables between these projected coherent states can be
computed in the limit of small amplitude, longwavelength density
fluctuations, which we refer to as the hydrodynamic limit. We then
discuss the relation between the charge and the topological charge
densities. Our approach is related to previous approaches of both 
that of Ezawa \cite{ezawa} and to that of Murthy and Shankar \cite{gmrs}.
Our coherent state wavefunctions are of the same form as in reference
\cite{ezawa} and our calculations are done in the same physical regime
(the hydrodynamic limit) as in reference \cite{gmrs}.

The issue of the classical limit is dealt with in section \ref{gibos}. We
show (in the hydrodynamic limit), that the projected coherent states 
satisfy the properties of generalized coherent states \cite{perel}. We also 
show that the off diagonal matrix elements of the observables vanish in
this limit. The theory thus admits a classical description in this
limit. Further, it can be written in completely in terms of bosonic fields
corresponding to gauge invariant physical observables.

The concluding section \ref{conc} summarises and discusses our results.

\section{Composite Boson theory and Coherent states}
\label{cbtcs}

In this section, we do the usual bosonization wherein the electronic theory  
is written in terms of bosons attatched to fluxes of Chern-Simons fields.
The electron operators and the procedure of flux attatchment is very
transparent in terms of coherent states. So we first define the Hilbert
space of the composite bosons and construct the coherent state basis for
it. We then construct the electron operators and obtain the
explicit mapping between the electronic Hilbert space and observables
and the gauge invariant states and observables in the bosonic theory.

\subsection{The Composite Boson Hilbert Space}
\label{hs}

The bosonic degrees of freedom are described by the spinor field operators
${\hat \varphi_\sigma(x)},~{\hat \varphi_\sigma^\dagger(x)},~\sigma=1,2 $ 
and the Chern-Simons gauge fields, $a_i(x),~i=1,2$. The $\hat \varphi$ 
operators act on the Hilbert space, $\cal H_B$ and satisfy the canonical
commutation relations,
\begin{eqnarray}
\left[{\hat \varphi}_{\sigma}(x),{\hat \varphi^\dagger}_{\sigma'}(y)\right] 
&=& \delta_{\sigma \sigma'}\delta^2(x-y) \nonumber \\
\label{ccrphi}
\left[{\hat \varphi}_{\sigma}(x),{\hat \varphi}_{\sigma'}(y)\right] & = & 
\left[{\hat \varphi}^{\dagger}_{\sigma}(x),{\hat
\varphi}^{\dagger}_{\sigma'}(y)\right] = 0
\end{eqnarray}
The Chern-Simons gauge fields act on the Hilbert space ${\cal H}_{CS}$ and 
satisfy, 
\begin{equation}
\label{ccrar}
\left[a_i(x),a_j(y)\right]~=~{\sqrt {\hbar c \over
\kappa}}\epsilon_{ij}\delta^2(x-y)
\end{equation}
where $\kappa = \frac{e^2}{2\pi \hbar c (2n+1)}$.
If we define the complex fields, $a(x)$ and $\bar a(x)$ as,
\begin{eqnarray}
a(x) &\equiv& \frac {a_2(x) + {\it i}a_1(x)}
{\sqrt 2} \sqrt \frac {\kappa}{\hbar c} \nonumber \\
{\bar a}(x) &\equiv& \frac {a_2(x) - {\it i}a_1(x)}{\sqrt 2} \sqrt
\frac {\kappa}{\hbar c}
\end{eqnarray} 
then the commutation relation between them is given by,
\begin{equation}
\label{ccrac}
\left[a(x),{\bar a}(y)\right] = \delta^2(x-y)
\end{equation}
The full Hilbert space of the composite boson theory is the direct sum
of the above two spaces and we denote it by,
\begin{equation}
{\cal H}_{\it CB}={\cal H}_{B} \oplus {\cal H}_{CS} 
\end{equation}
We denote the gauge invariant sector of this space by ${\cal H}_{\it phy} 
\subset {\cal H}_{CB}$. ${\cal H}_{\it phy}$ consists of the states which  
which respect the Chern-Simons Gauss law constraint, 
\begin{equation}
{\hat G}(x)|\psi \rangle_{\it {phy}} = 0
\end{equation}
where ${\hat G}(x)$ are the generators of gauge transformations given
by,
\begin{equation}
\label{gaucon}
{\hat G}(x)~=~\kappa \nabla \times {\vec a}(x) - 
{\it e}{\hat \varphi}^\dagger (x) {\hat \varphi}(x)
\end{equation}
We will refer to the gauge invariant observables, the operators that commute
with ${\hat G(x)}$ as physical obsevables.

\subsection{Coherent State Basis}
\label{csb}

In this section, we construct the coherent state basis for ${\cal H}_{CB}$.
The displacement operators are defined to be,
\begin{eqnarray}
 D(\alpha) &\equiv& e^{\int_x[\alpha(x){\bar a}(x) - {\bar 
\alpha}(x)a(x)]} \nonumber \\
U(\varphi) &\equiv& e^{\int_x[\varphi(x) {\hat \varphi}(x)^\dagger
-{\bar\varphi}(x){\hat\varphi}(x)]}
\end{eqnarray}
where, $\alpha(x) \equiv \frac {\alpha_2(x) + {\it 
i}\alpha_1(x)}{\sqrt 2} \sqrt \frac {\kappa}{\hbar c}$. 
The coherent states  $|\alpha, \varphi\rangle$,  parameterised by  the
gauge field $\alpha(x)$ and the spinor field $\varphi(x)$ are then given
by,
\begin{equation}
\label{cohdef}  
|\alpha, \varphi \rangle \equiv U(\varphi)D(\alpha)|0\rangle  \\
\end{equation}
where, 
\begin{equation}
\label{cbvac}
a(x)|0\rangle~=~{\hat \varphi}_\sigma (x) \vert 0 \rangle~=~0 
\end{equation}

The states defined in equation(\ref{cohdef}) can be intepreted as gaussian wave
packets peaked around the classical field configuration $(\alpha(x),
\varphi(x))$. They satisfy the three standard properties of coherent states 
\cite{perel} namely,

\noindent
1. Resolution of unity:
\begin{equation}
\label{rou}
\int {\cal D}[\alpha, \varphi]~|\alpha, \varphi \rangle
\langle\alpha, \varphi|~=~I 
\end{equation}
where ${\cal D}[\alpha, \varphi ]~=~\prod_{x,\sigma}
\frac{d\alpha(x) d{\bar\alpha}(x)}{2\pi {\it i}} 
\frac{d\varphi_\sigma (x) d{\bar\varphi}_\sigma (x)}{2\pi {\it i}}$

\noindent
2. Continuity of overlaps:
\begin{eqnarray}
\label{colap}
 \langle\alpha_1, \varphi_1 |\alpha_2, \varphi_2 \rangle~&=&  
e^{-\frac{{\it i}}{2}\frac {\kappa}{\hbar c}\int_x 
{\vec \alpha}_1(x) \times {\vec \alpha}_2(x) } e^{-\frac{{1}}{4}\frac 
{\kappa}{\hbar c}\int_x ({\vec \alpha}_1(x) - {\vec \alpha}_2(x))^2 }
\nonumber \\
 && .e^{\frac{1}{2}\int_x[{\bar\varphi}_1(x)  
\varphi_2(x) -\varphi_1(x){\bar\varphi_2}(x)]} 
e^{-\frac{1}{2}\int_x|\varphi_1(x)- \varphi_2(x)|^2} 
\end{eqnarray}

\noindent
3. Values of Observables:
\begin{equation}
\label{vobs}
\langle \alpha, \varphi| :O(a,{\bar a},{\hat \varphi},{\hat
\varphi}^\dagger): |\alpha, \varphi \rangle~=~
O(\alpha,{\bar \alpha}, \varphi, \varphi^\dagger)
\end{equation}

The coherent states are not gauge invariant. Under gauge
transformations,
\begin{eqnarray}
|\alpha,\varphi\rangle &\rightarrow& 
e^{\frac{i}{\hbar c}\int {\hat G}(x)\Omega(x)}|\alpha,\varphi\rangle \nonumber \\
&=& e^{\frac{{\it i}}{2}\frac {\kappa}{\hbar c}\int_x 
{\vec \alpha}(x) \times {\nabla \Omega}(x) } |\alpha-\nabla\Omega,\varphi
e^ {-\frac {{\it i}e}{\hbar c} \Omega}\rangle
\end{eqnarray}

We will now constuct a projection operator that projects any state into
the gauge invariant subspace, ${\cal H}_{\it phy}$. Consider,
\begin{equation}
\label{prop}
 P \equiv \frac {1}{V_G}\int_{\Omega} 
e^{ \frac{i}{\hbar c} \int_x \Omega(x){\hat G}(x)}
\end{equation}
where ${\hat G(x)}$ is the generator of gauge transformations, as given
in equation(\ref{gaucon}) and $V_G = \int_{\Omega}$, is the volume of the 
gauge group. Shifing the integration variable $\Omega$ by $\beta$ in the 
projection operator $P$,
\begin{equation} 
e^{\frac{i}{\hbar c} \int_x \beta(x){\hat G(x)}}P = P 
\end{equation} 
Taking $\beta \rightarrow 0$,
\begin{equation}
{\hat G(x)}P=0~~\Rightarrow~~{\hat G(x)}P|\psi\rangle = 0
\end{equation} 
This proves that $P$ is an operator that projects any state into ${\cal
H}_{\it phy}$.

The above three properties (\ref{rou} - \ref{vobs}) and the projection 
operator defined in equation(\ref{prop}) can be used to derive
the path integral representation of the gauge invariant evolution operator. 
This is done in the Appendix \ref{pir} to obtain,  
\begin{equation}
Z = \int {\cal D}[a_0(x,t)] {\cal D}[a_i(x,t)] {\cal D}[\varphi(x,t)] 
e^{\frac {i}{\hbar} \int dt d^2x {\cal L}(x,t)}  
\end{equation}
where ${\cal L}(x,t)$ is the standard lagrangian of matter fields
coupled to Chern-Simons gauge fields. This confirms the equivalence
of our formalism to the standard lagrangian formalism.

\subsection{Bosonization}
\label{bos}

We will now construct gauge invariant anticommuting operators that create 
and annihilate flux carrying bosons. These operators satisfy the fermionic 
canonical anticommutation relations and can hence be used to represent the
electron creation and annihilation operators in ${\cal H}_{CB}$. We will
thus be able to map the gauge invariant sector of the composite boson
Hilbert space, ${\cal H}_{\it phy}$, to the Hilbert space of the
electronic system, ${\cal H}_{el}$. The mapping is then used to map the
observables of the electronic system to gauge invariant operators in
${\cal H}_{CB}$.

 We define $c^{\dagger}_{\sigma}(x)$ as 
\begin{equation}
\label{elop}
  c^{\dagger}_{\sigma}(x) \equiv D(x) 
{\hat \varphi}^{\dagger}_{\sigma}(x) K(x)
\end{equation}
We have used $D(x)$ as short notation for $D(\alpha^v_x)$. 
$\alpha^v_x$ is the classical configuration of a vortex with a delta function
flux density at the point $x$.
\begin{equation}
\kappa \nabla \times {\vec \alpha^v_x}(z) = {\it e} \delta^2(z-x)
\end{equation}
$D(x)$ therefore creates a gaussian wave packet
peaked around this classical vortex configuration. When
$c^\dagger(x)$ acts on a state, ${\hat \varphi}^\dagger_\sigma (x)$
creates a bosonic particle at $x$ and $D(x)$ attatches Chern-Simons flux
to it. The operator $K(x)$ gives the Aharanov-Bohm phase corresponding
to all the other particles already present in the state. It is defined
as,
\begin{equation}
 K(x) \equiv e^{i(2n+1)\int_z \theta(x-z){{\hat \varphi}^{\dagger}(z)}
{\hat \varphi}(z)}
\end{equation}
where $\theta(x)$ is the angle the vector, $x$, makes with the x-axis. 

Using the commutation relations given in equations(\ref{ccrphi} and 
\ref{ccrar}), it can be verified that the following canonical anti-commutation 
relations hold good.
\begin{equation}
\{ c_{\sigma}(x), c_{\sigma'}^{\dagger}(y) \}~=~
\delta_{\sigma \sigma'}\delta^2(x-y)
\end{equation}
\begin{equation}
\{ c_{\sigma}(x), c_{\sigma'}(y) \} = 
\{ c_{\sigma}^{\dagger}(x),c_{\sigma'}^{\dagger}(y) \} = 0
\end{equation}
Hence  $c_{\sigma}^{\dagger} (x)$ and $c_{\sigma} (x)$
provide a representation of the electron creation and annihilation operators
in ${\cal H}_{CB}$. 

Under gauge transformations,
\begin{eqnarray}
{\hat \varphi}_\sigma (x) \rightarrow e^{i\frac{e}{\hbar c}\Omega (x)}
{\hat \varphi}_\sigma (x)~~&~,~&{\hat \varphi}^{\dagger}_\sigma (x)
\rightarrow e^{-i\frac{e}{\hbar c}\Omega (x)}{\hat \varphi}^{\dagger}_\sigma (x)
\nonumber \\
a_i(x) \rightarrow a_i(x)+\partial_i \Omega(x) &,&
~~D(x)\rightarrow e^{i\frac{e}{\hbar c}\Omega(x)}D(x)
\end{eqnarray}
We see that $c_\sigma(x)$ and $c^\dagger_\sigma (x)$ are gauge 
invariant. 

We are now in a position to map ${\cal H}_{el}$ into ${\cal H}_{\it
phy}$. We map the state with 0 number of electrons, $|0\rangle_{el}$,
to the vacuum state of ${\cal H}_{CB}$, defined in equation(\ref{cbvac}), 
projected to ${\cal H}_{\it phy}$.
\begin{equation}
\label{el0}
|0>_{el}~\rightarrow~P|0\rangle
\end{equation}
Since $c_\sigma(x)$ are gauge invariant, they commute with $P$. Then
from equation(\ref{cbvac}) it follows that,
\begin{equation}
c_\sigma (x)P|0\rangle ~=~ 0
\end{equation}
The state with $N$ electrons at $(x_1,x_2,....x_N)$ with spins $(\sigma_1,
\sigma_2,...,\sigma_N)$, $|\{x_n,\sigma_n\}_N\rangle$, is then mapped
into,
\begin{equation}
\label{eln}
|\{x_n,\sigma_n\}_N\rangle ~\rightarrow~ 
\prod_{n=1}^N c^\dagger_{\sigma_n}(x_n)P|0\rangle
\end{equation}
Since the states in the RHS of equations(\ref{el0}) and (\ref{eln}) 
form a basis for ${\cal H}_{el}$, these equations specify the explicit
mapping of ${\cal H}_{el}$ into ${\cal H}_{\it phy}$.

It is now easy to map the observables as well. The density is given by,
\begin{equation}
{\hat \rho}(x) =  c_{\sigma}^{\dagger}(x) c_{\sigma}(x) 
={\hat \varphi}_{\sigma}^{\dagger}(x) {\hat \varphi}_{\sigma}(x) 
\end{equation}
The spin density is,
\begin{equation}
{\hat S}^a(x) = \frac{1}{2}c_{\sigma}^{\dagger}(x) \tau_{\sigma {\sigma'}}
c_{\sigma'}(x) 
=\frac {1}{2} {\hat \varphi}_{\sigma}^{\dagger}(x) \tau_{\sigma {\sigma'}}
{\hat \varphi}_{\sigma}(x) 
\end{equation}

The current density is,
\begin{eqnarray}
{\hat J}_i(x) &=& \frac{1}{2}(c_{\sigma}^{\dagger}(x)
[-{\it i} \hbar \partial_{i} - \frac {e}{c}A_{i}(x)]c_{\sigma}(x)
+ h.c ) \nonumber \\
&=&\frac{1}{2}({\hat \varphi}_{\sigma}^{\dagger}(x) 
[-{\it i}\hbar\partial_{i}-\frac {e}{c}a_{i}(x)-\frac {e}{c}A_{i}(x)] 
{\hat \varphi}_{\sigma}(x)-  \nonumber \\
& &\frac{1}{c}{\hat \varphi}_{\sigma}^{\dagger}(x) {\hat \varphi}_{\sigma}(x) \int_z 
\alpha^v_{xi}(z){\hat G}(z) + h.c)
\end{eqnarray}
The last term acting on physical states is zero. Thus for matrix
elements between physical states, we have,
\begin{equation}
{\hat J}_{i}(x) = \frac{1}{2} {\hat \varphi}_{\sigma}^{\dagger}(x)
[-{\it i} \hbar \partial_{i} - \frac {e}{c}a_{i}(x) - \frac {e}{c}A_{i}(x)]
{\hat \varphi}_{\sigma}(x) + {\it h.c}
\end{equation}
Similarly, the kinetic energy density, ${\hat {\cal T}}(x) $ is
computed to be,
\begin{equation}
{\hat {\cal T}}(x) = \frac{1}{2m} {\hat \varphi}_{\sigma}^{\dagger}(x)
[-{\it i} \hbar \partial_{i} - \frac {e}{c}a_{i}(x) - \frac {e}{c}A_{i}(x)]^2
{\hat \varphi}_{\sigma}(x) 
\end{equation}

\section{Gauge Invariant Basis States}
\label{gibs}

In this section, we study the coherent states projected into ${\cal
H}_{\it phy}$. We show that these states form a basis of ${\cal H}_{\it
phy}$. Their wavefunctions and expectation values of observables are
computed. The LLL condition can then be seen to be equivalent to an
analyticity condition on the parameters. We then discuss the relation
between the charge density and the topological charge density. Finally,
we describe the parametrisation of the projected coherent states in
terms of a single complex spinor field $W_\sigma(x)$ discussed in the
end of section \ref{intro} and derive expressions for the observables in
terms of $W_\sigma(x)$

\subsection{Projected coherent states}
\label{pcs}
Consider the set of coherent states, projected to ${\cal H}_{\it phy}$,
\begin{equation}
|\alpha,\varphi\rangle_p \equiv P |\alpha,\varphi\rangle
\end{equation}
Using the fact that $P^2=P$ and equation(\ref{rou}), we have,
\begin{equation}
\label{roupcs}
\int {\cal D}[\alpha]{\cal D}[\varphi^\dagger]{\cal D}[\varphi]~
|\alpha,\varphi\rangle_p {_{p}} \langle \alpha,\varphi|~=~P
\end{equation}
$P$ is the identity operator in ${\cal H}_{\it phy}$ so the projected
coherent states form a basis for it.

The coherent states are not eigenstates of the number operator. Thus
they have a non-zero overlap with states containing any number of
particles. The wavefunction in the $N$ particle sector is the overlap
with the states given in equation(\ref{eln}),
\begin{equation}
\label{wf1}
\psi_N(\{x_i,\sigma_i\})~=~
\langle\{x_i,\sigma_i\}_N|\alpha,\varphi\rangle_p
\end{equation}
Using equations(\ref{cohdef}), (\ref{prop}), (\ref{elop}) and
(\ref{eln}), the RHS can be written as,
\begin{eqnarray}   
\psi_N(\{x_i,\sigma_i\})&=& \prod_{i>j}e^{i(2n+1)\theta(x_i - x_j)} \frac{1}{V_G} 
\int_{\Omega}e^{\frac{{\it i}}{2}\frac 
{\kappa}{\hbar c}\int_x {\Omega(x) (\nabla\times {\vec \alpha}(x)}) }
\prod_{i=1}^{N} \varphi_{\sigma_i}(x_i)
e^ {-\frac {{\it i}e}{\hbar c} \sum_{i=1}^{N}\Omega(x_i)}
e^{-\frac{1}{2}\int_x |\varphi(x)|^{2}} \nonumber \\ & &
 e^{-\frac{{\it i}}{2}\frac {\kappa}{\hbar c}\int_x 
(\sum_{i=1}^{N}{\vec \alpha_i}(x)) \times ({\vec \alpha}(x)  - \nabla\Omega(x))}  
e^{-\frac{{1}}{4}\frac {\kappa}{\hbar c}\int_x ({\vec \alpha}(x) - \nabla\Omega(x) - 
\sum_{i=1}^{N}{\vec \alpha_i}(x))^2 } \label{wf2}
\end{eqnarray} 
The details of this calculation and what follows is given in
Appendix \ref{wfcal}. The $\Omega$ integral in the RHS above is 
gaussian and can be done exactly. After some algebra, we obtain,
\begin{eqnarray}
\psi_N(\{x_i,\sigma_i\})&=& {\it const}.
e^{-\frac{1}{2}\int_x |\varphi(x)|^{2}}
e^{-\frac{{1}}{2}\frac {\kappa}{\hbar c}
\int_x {\nabla \Omega_T(x) \cdot \nabla \Omega_T(x)} } 
\nonumber \\
&&\times \prod_{i=1}^{N}[ \varphi_{\sigma_i}(x_i)
e^{\frac{e}{\hbar c} 
\{\Omega_T(x_i)-\bar{\Omega}_T(x_i) - i\Omega_L(x_i)\} } ] 
\psi_{L}(\{x_i\}) 
\label{wf3}
\end{eqnarray}
where $\psi_L(\{x_i\})$ is the Laughlin wavefunction, 
\begin{equation}
\psi_{L}(\{x_i\}) = \prod_{i > j} (z_i - z_j)^{2n+1} 
e^{-\frac {1}{4l_c^2} \sum_i |z_i|^2}
\end{equation}
$\alpha$ has been written as 
\begin{equation}
\alpha_i(x)~=~\epsilon_{ij}\partial_j \Omega_T(x)+\partial_i \Omega_L(x)
\end{equation}
and $\bar{\Omega}_T(x)= -{\frac {\hbar c}{e}}{\frac {|x|^2}{4{l_c}^2}}$.
Note that when $\varphi_\sigma(x)=constant$ and $\Omega_T(x) = {\bar
\Omega}_T(x)~\Rightarrow~\nabla \times {\vec \alpha}=B$, the
wavefunction reduces to the Laughlin wavefunction. Thus the "mean
field" state is the Laughlin state. In this case the $\Omega$
integral in equation(\ref{wf2}) is equivalent to an $N$ vertex
operator correlation function in a c=1 conformal field theory. These
wavefunctions are exactly of the form written down by Ezawa
\cite{ezawa}.

\subsection{Parameterisation and LLL condition}
\label{plll}

Apart from an overall factor that only affects the norm, the wavefunction
in equation(\ref{wf3}) depends on the parameters $\alpha$ and
$\varphi$ through a spinor field $W_\sigma(x)$ defined as,
\begin{equation}
\label{wdef}
W_\sigma(x)~\equiv~ \varphi_{\sigma_i}(x_i)
e^{\frac{e}{\hbar c}
(\Omega_T(x_i)-\bar{\Omega}_T(x_i)-i\Omega_L(x_i)) }  
\end{equation}
$W_\sigma(x)$ and hence the wavefunction is gauge invariant (as it
should be), since under gauge tranformations,
\begin{eqnarray}
\Omega_T(x)&~\rightarrow~&\Omega_T(x)           \nonumber \\
\Omega_L(x)&~\rightarrow~&\Omega_L(x)+\Omega(x) \nonumber \\
\varphi_\sigma(x)&~\rightarrow~&\varphi_\sigma(x)
e^{\frac{ie}{\hbar c}\Omega(x)}
\end{eqnarray}
$\alpha$ and $\varphi$ have 6 real field components. The gauge
invariance of the wavefunctions reduces the number of parameters to
5. There is another local invariance of $W$, i.e.
\begin{eqnarray}
\Omega_T(x)&~\rightarrow~&\Omega_T(x)+\chi(x)   \nonumber \\
\Omega_L(x)&~\rightarrow~&\Omega_L(x)           \nonumber \\
\varphi_\sigma(x)&~\rightarrow~&\varphi_\sigma(x)
e^{-\frac{e}{\hbar c}\chi(x)}
\end{eqnarray}
Only the norm of the state changes under this transformation and
the physical state remains the same. Clearly this transformation is
not unitarily implemented in ${\cal H}_{CB}$. Nevertheless it reduces
the number of independent real fields that parameterize the states to
4, the components of the spinor field $W$. Thus we can define the
normalised projected coherent states, that are parameterised by $W$
as,
\begin{equation}
\label{wstdef}
|W\rangle~=~\frac{1}{\cal N}|\alpha,\varphi\rangle_p
\end{equation}
where, ${\cal N}~=~{_p}\langle \alpha,\varphi|\alpha,\varphi\rangle_p$, 
is the norm of $|\alpha,\varphi\rangle_p$. 

From equations (\ref{wf3}) and (\ref{wdef}) it is clear that the LLL
condition is equivalent to the condition that $W$ is analytic,
\begin{equation}
\label{lllcon}
\partial_{\bar z} W_\sigma(x)~=~0
\end{equation}
Thus the LLL condition is easily implemented in this formalism as it
is equivalent to an analyticity condition on the parameters.

\subsection{Observables}
\label{obs}

We will now compute the expectation values of gauge invariant 
operators in the projected coherent states. This is given by,
\begin{equation}
\langle {\hat O} \rangle~=~\langle W |{\hat O}| W \rangle
\end{equation}
where ${\hat O}$ is a gauge invariant observable. 

We do all our calculations the limit of $W_\sigma(x)$ being a slowly
varying function of $x$ (over a length scale of $l_c$). As we will
see, this is also the limit of small density fluctuations. We refer
to this limit as the hydrodynamic limit. We note that this is also
the limit in which the analytic calculations of Murthy and Shankar
\cite{gmrs} are done. 

Just as in the case of the Laughlin wavefunction, the computation of 
${\cal N}$ reduces to the computation of the partition function of a
classical 2-d plasma problem. Except that here, the plasma density is 
coupled to an external field which is a function of $W$. In the 
hydrodynamic limit, the partition function can be evaluated by the saddle
point approximation. The details of the calculation are presented in
Appendix \ref{obscal}, where we evaluate the norm to be,
\begin{eqnarray}
{\cal N}[W,\Omega_T] &=& const \times
e^{-\int_x  W^\dagger(x)W(x) e^{-\frac{2e}{\hbar c} 
\{\Omega_T(x)-\bar{\Omega}_T(x) \}}} 
e^{-\frac{\kappa}{\hbar c}\int_x 
\nabla \Omega_T(x) \cdot \nabla \Omega_T(x)} 
\nonumber \\
&&e^{-\frac{1}{8\pi(2n+1)}\int_x 
[\ln (W^\dagger(x)W(x))+2\frac{e}{\hbar c} \bar{\Omega}_T(x)]
\nabla^2 [\ln (W^\dagger(x)W(x))+2\frac{e}{\hbar c} \bar{\Omega}_T(x)]}
\label{normexp}
\end{eqnarray}
Expectation values of observables similarly reduce to the computation
of expectation values in the plasma problem. These are also computed
in the saddle point approximation in Appendix \ref{obscal}. We get
the density to be,
\begin{eqnarray}
\rho(x)-{\bar \rho} &\equiv&  \langle W|{\hat \rho}(x)-{\bar \rho}|W\rangle 
\nonumber \\
&=& \frac {-1}{4\pi(2n+1)} 
\nabla^2 \ln (W^\dagger(x)W(x))
\label{rho}
\end{eqnarray}
where $\bar \rho$ is the mean density.

Similarly, the spin density is computed to be,
\begin{eqnarray}
s^a(x) &\equiv&  \langle W|{\hat s}^a(x)|W\rangle \nonumber \\
&=& \frac {\rho (x)}{2} Z^\dagger(x)\tau^a Z(x)
\label{spin}
\end{eqnarray}
where we have denoted the normalised spinor by $Z$,
\begin{equation}
\label{zdef}
Z_\sigma(x) \equiv \frac {W_\sigma(x)}{\sqrt{W^\dagger(x)W(x)}}
\end{equation}
and $\tau^a$ are the Pauli spin matrices.
The current density and the kinetic energy density are also 
computed to yield,
\begin{eqnarray}
J_i(x) &\equiv&  \langle W|{\hat J}_i(x)|W\rangle \nonumber \\
&=& \rho(x)[\hbar L_i^3(x) - \frac{e}{c}(A_i(x)-\alpha_i(x))]
\end{eqnarray}
where $L^3_i \equiv \frac{1}{2i}(Z^\dagger \partial_i Z-h.c)$ and 
$\kappa \nabla \times {\vec \alpha}(x) = {\it e} \rho(x)$,
\begin{eqnarray}
{\cal T}(x) &\equiv&  \langle W|{\hat {\cal T}}(x)|W\rangle 
\nonumber \\
&=&
\hbar \omega_c \rho(x) 
\frac {\partial_z\bar{W}_{\sigma}(x) \partial_{\bar{z}}W_{\sigma}(x)}
{ \bar{W}_{\sigma'}(x)W_{\sigma'}(x) } 
\end{eqnarray}
Note that the kinetic energy density is zero when $W$ is analytic.

\subsection{Charge and Topological Charge Densities}
\label{ctcd}

The topological charge density is given by
\begin{equation}
\label{top1}
q(x)~=~\frac{1}{8\pi}\epsilon_{ij}
{\hat n}(x)\cdot\partial_i{\hat n}(x)\times\partial_j{\hat n}(x)
\end{equation}
where ${\hat n}(x)$ is the local direction of spin polarization,
${\vec s}(x)=\frac{1}{2}\rho(x){\hat n}(x)$. In terms of $Z$, it is 
given by,
\begin{equation}
\label{top2}
q(x)~=~\frac{1}{2\pi i}\epsilon_{ij}\partial_iZ^\dagger(x)\partial_jZ(x)
\end{equation}
As can be seen from equations (\ref{rho}) and (\ref{top2}), the
topological charge density is not necessarily proportional to the 
electrical charge density. In fact, in general, they are independent
of each other since $W^\dagger(x)W(x)$ and $Z_\sigma(x)$ are
independent variables. However, if the LLL condition is satisfied, 
then the analyticity of $W_\sigma(x)$ relates the modulus and the 
phase of each component. Then $W^\dagger(x)W(x)$ and $Z_\sigma(x)$ 
are no longer independent. In fact if we use the analyticity condition,
$\partial_i W_\sigma (x)~=~-i\epsilon_{ij}\partial_jW(x)$, in the RHS of
equation (\ref{rho}), we get,
\begin{equation}
\label{eltop}
\rho(x)-\bar{\rho}= -\frac{1}{2n+1} q(x)
\end{equation}
Thus the topological charge density is proportional to the electrical
charge density if and only if the LLL condition is satisfied. The relation 
(\ref{eltop}) will therefore not be true in presence of Landau level
mixing. 

When the densities are proportional, the total excess charge, $Q$, will of
course be proportional to the total topological charge, $Q_{top}$. However,
the total charges could be proportional without the densities being so.
We will now investigate this possibility. Integrating equation (\ref{rho}) 
over all space, we have,
\begin{equation}
\label{qtot1}
Q~=~\frac{1}{4\pi(2n+1)}\oint dx^i \epsilon_{ij}
\partial_j \ln(W^\dagger(x)W(x))
\end{equation}
where the contour is at infinity. If $W$ is analytic at infinity, then
the RHS of equation (\ref{qtot1}) can be written as, 
\begin{eqnarray}
Q&~=~&-\frac{1}{2\pi(2n+1)}\oint dx^i \frac{1}{2i}
(Z^\dagger(x)\partial_iZ(x)-\partial_iZ^\dagger(x) Z(x)) \nonumber \\
&~=~& -\frac{1}{2n+1}Q_{top}
\label{qtot2}
\end{eqnarray}
Thus if there is no Landau level mixing in the ground state, the total
charge is always proportional to the topological charge. Note that $Z$
and hence $q(x)$ is well defined only if $\rho(x)$ is non-zero
everwhere. So all our considerations are true only in this case. They
will not hold for polarized vortices where $\rho(x)$ will vanish at some
point.

\section{Gauge Invariant Bosonization}
\label{gibos}
In the previous section, we saw that the projected coherent states are 
labelled by a spinor field $W$, and that the expection values of observables
could be computed in the hydrodynamic limit in terms of $W$. The states can 
therefore be labelled by the values of the physical observables, the density
$\rho(x)$ and the normalised spinor $Z_\sigma(x)$. In this section, 
we will show that these states themselves satisfy the generalised coherent 
state properties \cite{perel} in ${\cal H}_{phy}$. Namely, the resolution of 
unity and continuity of overlaps. This implies that, in the hydrodynamic
limit, the original electronic theory can be expressed completely
in terms of bosonic field operators corresponding to $\rho(x)$ and 
$Z_\sigma(x)$. Thus in this limit, the theory can be bosonized in a gauge 
invariant way with no redundant degrees of freedom. 

\subsection{Resolution of Unity}
\label{ru}

The fact that the identity operator in ${\cal H}_{phy}$ can be resolved in 
terms of the projected coherent states has already been shown in
equation (\ref{roupcs}). Here we express this same equation in terms of
the gauge invariant parameters. We perform the following change of variables
in equation (\ref{roupcs}),
\begin{equation}
\label{cov}
\alpha_i(x), \varphi_\sigma(x)~\rightarrow~
\Omega_L(x),\Omega_T(x), W_\sigma(x)
\end{equation}
further, using equations (\ref{wstdef}) and (\ref{normexp}), we get
\begin{eqnarray}
 I &~=~&
\int {\cal D}[\alpha, \varphi]~|\alpha, \varphi \rangle_{p~p}
\langle\alpha, \varphi| \nonumber \\
 &~=~& 
{\it const} \int {\cal D}[\Omega_T(x)]{\cal D}[\Omega_L(x)]{\cal D}[W] 
e^{-\int_x \frac{4e}{\hbar c} \Omega_T(x)}
   {\cal N}[W,\Omega_T] |W \rangle \langle W|  \nonumber \\
&~=~& {\it const} \int {\cal D}[W] {\cal G}[W] |W \rangle \langle W| 
\label{rouw}
\end{eqnarray}
where the factor $e^{- \frac{4e}{\hbar c}\Omega_T(x)}$
is the Jacobian due to the change of variables $\varphi \rightarrow W$,
\begin{equation}
\label{dw} 
{\cal D}[W]=\prod_{x,\sigma}
\frac{d W_\sigma (x) d{\bar W}_\sigma (x)}{2\pi {\it i}}
\end{equation}
and
\begin{equation}
\label{gdef}
{\cal G}[W]\equiv 
\int{\cal D}[\Omega_T] e^{-\int_x \frac{4e}{\hbar c}\Omega_T(x)} 
{\cal N}[W,\Omega_T]
\end{equation}
The integral over $\Omega_T$ can be done in the hydrodynamic limit, the
details are given in Appendix \ref{omegatint}, to get,
\begin{equation}
\label{gexp}
{\cal G}[W] ~=~ 
{\it const} \prod_x \frac {1}{[W^\dagger(x)W(x)]^2}
\end{equation}
We now make another change of variables from $W_\sigma(x)$ to 
$\rho(x), Z_\sigma(x)$, defined by equations (\ref{rho}) and
(\ref{zdef}), to get,
\begin{equation}
\label{rouz}
I~=~const~\int {\cal D}[\rho]{\cal D}[Z]~
|\rho,Z\rangle\langle\rho,Z|
\end{equation}
where,
\begin{equation}
{\cal D}[\rho]~=~\prod_x d\rho(x)~~{\cal D}[Z]~=~\prod_x 
sin^2\theta(x)sin\phi(x)d\theta(x)d\phi(x)d\psi(x)
\end{equation}
$Z$ has been parameterised as 
$$Z=(cos\frac{\theta}{2}e^{i(\frac{\psi+\phi}{2}}),
sin\frac{\theta}{2}e^{i(\frac{\psi-\phi}{2})})$$

\subsection{Overlaps}
\label{olap}

The overlap of two gauge invariant coherent states 
$|W_1\rangle$ and $|W_2\rangle$ is computed in the hydrodynamic limit in
Appendix \ref{olapcal}. The final answer is,

\begin{eqnarray}
\langle W_1|W_2\rangle &~=~& e^{-F[\rho_1,Z_1,\rho_2,Z_2}]\nonumber \\
F &~=~&-\frac{i}{4}\int_x(\rho_1(x)+\rho_2(x))\Phi({\hat n}_1,{\hat n}_2)
\nonumber \\
&&-\frac{1}{4}\int_x(\rho_1(x)+\rho_2(x))
\ln (\frac{1+{\hat n}_1\cdot {\hat n}_2}{2}) 
\nonumber \\
&&-\frac{\pi}{2}(2n+1)\int_{x,y} 
(\rho_1(x)-\rho_2(x))\langle x|\frac{1}{\nabla^2}|y\rangle
(\rho_1(y)-\rho_2(y)) 
\label{olapexp}
\end{eqnarray}
where $\Phi({\hat n}_1,{\hat n}_2)$ is the solid angle subtended by the
geodesic triangle with ${\hat n}_1, {\hat n}_2$ and some third point on
the unit sphere as vertices.

Note that the overlap smoothly goes to 1 as $(\rho_1,Z_1)\rightarrow
(\rho_2,Z_2)$. We will now evaluate the overlap for neighbouring states.
We put $\rho_2(x) = \rho(x)$, $\rho_1(x) = \rho(x) + 
\epsilon \partial_{\it t}\rho(x)$, $Z_{2\sigma}(x) = Z_{\sigma}(x)$ 
and $Z_{1\sigma}(x) = Z_{\sigma}(x) + 
\epsilon \partial_{\it t} Z_{\sigma}(x)$. Keeping terms only to the 
order $O(\epsilon)$ we get,

\begin{equation}
\label{olapeps}
\langle W+ \epsilon \partial_{\it t}W|W\rangle 
= e^{-{\it i} \epsilon \int_x \rho(x)L_{{\it t}}^{3}(x)+O(\epsilon^2)}
\end{equation} 
where $L^3_{\mu} \equiv \frac{1}{2i}(Z^\dagger \partial_{\mu} Z-h.c)$ for 
$\mu = {\it t}$, $1$ and $2$. 

If we now impose the LLL condition, then the charge density fluctuations
get tied up to the the spin density fluctuations. i.e.,
\begin{eqnarray}
\rho(x)&~=~&\bar{\rho}- \frac{1}{2n+1} q(x) \nonumber \\
&~=~& {\bar \rho}-\frac {1}{2\pi(2n+1)} \epsilon_{ij} \partial_i L_j^3(x)
\end{eqnarray}
The theory can then be expessed in terms of spin fluctuations alone. The 
expresion for the overlap in equation (\ref{olapeps}) then gets written as,
\begin{equation}
\label{olaplll}
\langle W+ \epsilon \partial_{\it t}W|W\rangle =
e^{ -{\it i} \epsilon \bar{\rho} \int_x L_{{\it t}}^{3}(x) + 
{\it i} \epsilon \frac{1}{4\pi(2n+1)} \int_x \epsilon_{\mu \nu \lambda} 
L_{\mu}^{3}(x) \partial_{\nu} L_{\lambda}^{3}(x)  + O (\epsilon^2)}
\end{equation}
The second term in the exponent in the RHS of equation (\ref{olaplll})
is the Hopf term. Thus the theory, when restricted to the lowest Landau
level is a NLSM with a Hopf term in the action.

\section{The classical limit}
\label{cl}

We will finally show that for large skyrmions, the theory becomes
classical. Consider the set of states corresponding to configurations chaqracterised 
by a size parameter, $\lambda$. We parameterie them as,
\begin{eqnarray}
\rho^{\lambda}(x)&~=~& \bar{\rho} + \frac{1}{\lambda^2} 
\Delta \rho (\frac{x}{\lambda}) \nonumber \\
Z^{\lambda}(x)&~=~& Z(\frac{x}{\lambda})
\end{eqnarray}
Substituting $\rho_1^{\lambda}(x)$, $z_{1\sigma}^{\lambda}(x)$ 
and $\rho_2^{\lambda}(x)$, $z_{2\sigma}^{\lambda}(x)$ 
in equation (\ref{olapexp}) and changing the variable 
$x \rightarrow \lambda x$ we get,
\begin{equation}
\langle W_1|W_2\rangle = e^ 
{\int_x [\lambda^2 \bar{\rho}\ln (\frac{1+{\hat n}_1\cdot {\hat n}_2}{2}) 
+ 0(\lambda^0)] }
\end{equation}
$\frac{1}{2} \ln (\frac{1+{\hat n}_1\cdot{\hat n}_2}{2})$ is zero 
${\hat n}_1 = {\hat n}_2$ and negative otherwise.  Thus for $W_1 \neq W_2$, 
\begin{equation} 
\lim_{\lambda \rightarrow \infty} \langle W_1|W_2\rangle \rightarrow 0
\end{equation}
The coherent states thus become orthogonal when $\lambda \rightarrow 0$.
It can also be shown that the off-diagonal matrix elements of the
observables in the coherent state basis, vanish in this limit. Hence the
set of states corresponding to a system of skyrmions will behave
classically in the limit of the skyrmion sizes tending to infinity.

\section{Conclusion}
\label{conc}

The motivation of this work was to examine the microscopic basis of the
semiclassical NLSM for skyrmions at $\nu = 1/(2n+1)$. In the systems of 
interest, the energy scales are such that it is important to impose the 
LLL constraint. The coherent state basis is the ideal one to address 
questions of quantum states in a semiclassical approach. Therefore, we 
developed a coherent state formalism of the composite boson theory.

We showed that the coherent state basis of ${\cal H}_{CB}$, when
projected to the physical subspace ${\cal H}_{phy}$, can be
parameterized by a spinor field that we denoted by $W_\sigma(x)$. In the
hydrodynamic limit we have shown that these states, $|W\rangle$ themselves 
satisfy the coherent state properties of the resolution of unity and
continuity of overlaps. The LLL condition is equivalent to the condition
that $W_\sigma(x)$ are analytic functions. 

The charge density is determined by the modulus of $W$ i.e
$W^\dagger(x)W(x)$ and the spin density by the normalised $CP_1$ spinor,
$Z_\sigma(x)$. In general these are independent quantities and therefore
the charge density is independent of the spin density. However if
$W(x)$ is analytic, the modulus and phase of each of its components get
tied up. We showed, that consequently, the excess charge density becomes
proportional to the topological charge density which is determined by
the spin density. Thus this proportionality will cease to hold in
presence of Landau level mixing. We also showed that the condition for 
the total charge density to be proportional to the topological charge is
weaker. It only requires $W(x)$ to be analytic at infinity. i.e. that the
ground state does not have Landau level mixing. 

Finally we showed that if we consider the set of states corresponding to
classical configuations of characterised by a length scale $\lambda$,
then they become orthogonal in the limit of $\lambda \rightarrow
\infty$. This implies that a system of skyrmions will behave classically
in the limit of their sizes going to infinity.

\begin{appendix}

\section{The path integral representation}
\label{pir}

 In the following appendix we derive the path integral representation
of the partition function by 
splitting the time interval $t$ into $N$ segments of length $\epsilon$
and take the limit $\epsilon \rightarrow 0$, $N \rightarrow 
\infty$ such that $\epsilon N = t$. And at each intermediate step we insert 
the resolution of identity (\ref{roupcs}) of ${\cal H}_{\it phy}$ ,
\begin{equation}
  Z = {\it Tr}  e^{\frac {-i}{\hbar}Ht} =
     {\it Tr} [ e^{\frac {-i}{\hbar}\epsilon H}]^N  
\end{equation}
\begin{equation}
  Z = \prod_{n=0}^N \int_{\alpha_n,\varphi_n,\Omega_n}\prod_{n=0}^N
    \langle\alpha_{n+1}\varphi_{n+1}|Pe^{\frac {-i}{\hbar}H\epsilon}P
    |\alpha_n \varphi_n\rangle 
\end{equation}
   where $(\alpha_{N+1},\varphi_{N+1}) \equiv (\alpha_0,\varphi_0)$

Since $H$ commutes with $P$ and $P^2 = P$, we get, after explicitly acting
$P$ on $|\alpha_n \varphi_n\rangle$  and  making use of gauge invariance of 
$|\alpha\varphi\rangle\langle\alpha\varphi|$, we get
\begin{equation}
  Z = \prod_{n=0}^N \int_{\alpha_n,\varphi_n,\Omega_n} \prod_{n=0}^N t_n
\end{equation}
where,
\begin{equation}
  t_n = e^{\frac{{\it i}}{2}\frac {\kappa}{\hbar c}\int_x 
{\vec \alpha_n}(x) \times {\nabla \Omega_n}(x) } 
\langle\alpha_{n+1}-\nabla\beta_{n+1},\varphi_{n+1} e^ {-\frac {{\it 
i}e}{\hbar c} \beta_{n+1}}| e^{\frac {-i}{\hbar}\epsilon H}|
 \alpha_n-\nabla\Omega_n-\nabla\beta_n,\varphi_n
e^ {-\frac {{\it i}e}{\hbar c} (\Omega_n+\beta_n)}\rangle
\end{equation}
   To the order $\epsilon$: $\alpha_{n+1}=\alpha_n + \epsilon \dot{\alpha}_n$,
 $\Omega_{n+1}=\Omega_n + \epsilon \dot{\Omega}_n$ and 
 $\varphi_{n+1}=\varphi_n + \epsilon \dot{\varphi}_n$. And if we choose 
$\beta_{n+1} = \beta_n + \Omega_{n+1}$ then to the order $O(\epsilon)$ $t_n$ is:
\begin{eqnarray}
  t_n &=& e^{\frac{{\it i}}{2}\frac {\kappa}{\hbar c}\int_x 
{\vec \alpha_n}(x) \times {\nabla \Omega_n}(x) } \nonumber \\ 
& & \langle\alpha_n-\nabla\beta_n-\nabla\Omega_n + 
\epsilon(\dot{\alpha}_n - \nabla\dot{\Omega}_n),\{\varphi_n + 
\epsilon(\dot{\varphi}_n-\frac {i e}{\hbar c}\varphi_n\dot{\Omega}_n)\} e^ 
{-\frac {{\it i}e}{\hbar c} (\beta_n + \Omega_n)}| e^{\frac 
{-i}{\hbar}\epsilon H} \nonumber \\
& &      |\alpha_n-\nabla\beta_n-\nabla\Omega_n,\varphi_n
e^ {-\frac {{\it i}e}{\hbar c} (\Omega_n+\beta_n)}\rangle
\end{eqnarray}
  Using  the fact that  
$\epsilon \dot{\beta}_n =\beta_n-\beta_{n-1} + O(\epsilon^2)$ and
\begin{equation}
\langle\alpha+\delta\alpha,\varphi+\delta\varphi| e^{\frac {-{\it i}}{\hbar}\epsilon H}
| \alpha,\varphi\rangle =
\langle\alpha+\delta\alpha,\varphi+\delta\varphi|\alpha,\varphi\rangle 
[1 - {\it i} \frac {\epsilon}{\hbar} \langle\alpha\varphi|H|\alpha\varphi\rangle] + 
O(\epsilon^2) 
\end{equation}
where $\delta\alpha \sim O(\epsilon)$ and $\delta\varphi \sim O(\epsilon)$  
the above expression for $t_n$ , after defining 
$\alpha_{0n} \equiv \dot{\Omega}_n / c$, and making use of gauge invariance of
$H$, we get
\begin{eqnarray}
  t_n = \exp[\frac {i \epsilon}{\hbar}\int_x \{ - \frac {\kappa}{2} 
\epsilon_{\mu\nu\lambda} 
\alpha_n^{\mu}(x) \partial^{\nu} \alpha_n^{\lambda}(x) + 
 e \alpha_{0n}(x) \bar{\varphi}_n(x)\varphi_n(x) -\nonumber \\
{\it i}\hbar \frac {1}{2}
[\varphi_n(x) \dot{ \bar{\varphi}}_n(x) - \bar{\varphi}_n(x)
\dot {\varphi}_n(x)] - H(\varphi_n,\alpha_n)\}]   
\end{eqnarray}
    If we now take the limit $\epsilon \rightarrow 0$ the partition 
function becomes (after calling $\alpha$ by $a$) 
\begin{equation}
   Z = \int {\cal D}[a_0(x,t)] {\cal D}[a_i(x,t)] {\cal D}[\varphi(x,t)] 
e^{\frac {i}{\hbar} \int dt d^2x {\cal L}(x,t)}
\end{equation}
 where ${\cal L}(x,t)$ is the standard Chern-Simons lagrangian.

\section{The wavefunctions}
\label{wfcal}

   In this appendix we give the details of the calculation of 
equation(\ref{wf2}). Using equations (\ref{prop}), (\ref{elop}) and 
(\ref{eln}) we get,
\begin{equation}
\label{awf0}
  \psi_{N}(\{x_i,\sigma_i\})= \frac{1}{V_G} \int_{\Omega}e^{\frac{{\it 
i}}{2}\frac {\kappa}{\hbar c}
\int_x {{\vec \alpha}(x)} \times \nabla\Omega(x)} \langle 0 
|{\prod_{i=1}^{N}} 
[K(x_i)\hat{\varphi}_{\sigma_i}(x_i)D^{\dagger}(x_i)]
|\alpha-\nabla\Omega,\varphi
e^ {-\frac {{\it i}e}{\hbar c} \Omega}\rangle
\end{equation}
Using the fact that,
\begin{equation}
\hat{\varphi}_{\sigma_i}(x_i)K(x_j) = e^{i(2n+1)\theta(x_i - x_j)} 
K(x_j)\hat{\varphi}_{\sigma_i}(x_i)
\end{equation}
and $\langle|0|K(x)=0$, we can pull all the $K'$s to the left and rewrite
equation (\ref{awf0}) as,
\begin{eqnarray}   
\psi_{N}(\{x_i,\sigma_i\})&=& \prod_{i>j}e^{i(2n+1)\theta(x_i - x_j)} 
\frac{1}{V_G} \int_{\Omega}e^{\frac{{\it i}}{2}\frac 
{\kappa}{\hbar c}\int_x {\Omega(x) (\nabla\times {\vec \alpha}(x)}) }
\prod_{i=1}^{N} \varphi_{\sigma_i}(x_i)
e^ {-\frac {{\it i}e}{\hbar c} \sum_{i=1}^{N}\Omega(x_i)}
e^{-\frac{1}{2}\int_x |\varphi(x)|^{2}} \nonumber \\ & &
 e^{-\frac{{\it i}}{2}\frac {\kappa}{\hbar c}\int_x 
(\sum_{i=1}^{N}{\vec \alpha_i}^{v}(x)) \times ({\vec \alpha}(x) - \nabla\Omega(x))}  
e^{-\frac{{1}}{4}\frac {\kappa}{\hbar c}\int_x ({\vec \alpha}(x) - \nabla\Omega(x) - 
\sum_{i=1}^{N}{\vec \alpha_i}^{v}(x))^2 } \label{eq:wf1}
\end{eqnarray} 
This is equation (\ref{wf2}). We now write,
\begin{equation}
\alpha_m^{i}(x) = \epsilon ^{ij} \partial_j f_m(x)
\end{equation}
where,
\begin{equation}
\label{fdef}
-\kappa \nabla^2 f_m(x) = e \delta(x-x_m)
\end{equation}
for $m= 1-N$. 

The zero momentum mode of the $\Omega$ integral will make the wavefunction 
vanish unless the total number of flux quanta equals the total numberof 
particles. i.e.
\begin{equation}
\kappa \int_x \nabla \times  {\vec \alpha}(x) = e N
\end{equation}
The $\Omega$ integral is gaussian for the other modes and can be done exactly
to give,
\begin{equation}
{\it const} \times exp[{-\frac{{1}}{4}\frac {\kappa}{\hbar c}\int_x ({\vec 
\alpha}(x) - \sum_{i=1}^{N}{\vec \alpha_i}(x))^2 }]
\end{equation}

We then write the wavefunction as,
\begin{eqnarray}
\psi_N(\{x_i,\sigma_i\})&~=~&const\times e^{-\frac{1}{2}\int_x |\varphi(x)|^{2}}
\prod_{i>j}[e^{i\theta(x_i - x_j)}]^{2n+1} \nonumber \\
&&\prod_{i=1}^{N}[ \varphi_{\sigma_i}(x_i) e^{-i \frac{e}{hbar c} \Omega_L(x)}]
e^{-\frac{{1}}{2}\frac {\kappa}{\hbar c}\int_x ({\vec \alpha}(x) - 
\sum_{i=1}^{N}{\vec \alpha_i}(x))^2 } 
\end{eqnarray}
We also have,
\begin{eqnarray}
\label{alpha2}
\int_x {\vec \alpha}(x).{\vec \alpha_m}(x)&~=~&
\int_x {\nabla \Omega_T(x)}.{\nabla f_m(x)} \nonumber \\
&~=~&-\int_x  \Omega_T(x){\nabla}^2 f_m(x) \nonumber \\
&~=~&{\frac {e}{\kappa}} \Omega_T(x_m)
\end{eqnarray}
Using the fact that  the solution of equation (\ref{fdef}) is,
\begin{equation}
f_m(x) = -{\frac {e}{\kappa}}{\frac {1}{2\pi}} \ln {|x - x_m|}
\end{equation}
and proceeding as in equation (\ref{alpha2}), we have,
\begin{eqnarray}
\int_x {\vec \alpha_m}(x).{\vec \alpha_n}(x)&~=~&{\frac {e}{\kappa}} f_m(x_n) 
\nonumber \\
&~=~&-{\frac {e^2}{\kappa^2}}{\frac {1}{2\pi}} \ln {|x_m - x_n|}
\end{eqnarray} 
So we have the result,
\begin{eqnarray}
\int_x ({\vec \alpha}(x)-\sum_{i=1}^{N}{\vec \alpha_i}(x))^2 
&~=~&\int_x \nabla\Omega_T(x).\nabla\Omega_T(x)-
{\frac {e^2}{\kappa^2}}{\frac {1}{2\pi}} \sum_{m \neq n}^{N}\ln {|x_m - x_n|}
\nonumber \\
&&- 2{\frac {e}{\kappa}} \sum_{m=1}^{N}\Omega_T(x_m) + const 
\end{eqnarray}
The (infinite) $const$ comes from the $m=n$ terms 

Finally the wavefunction is written as,
\begin{eqnarray}
\psi_N(\{x_i,\sigma_i\})&~=~&const\times e^{-\frac{1}{2}\int_x |\varphi(x)|^{2}}
e^{-\frac{{1}}{2}\frac {\kappa}{\hbar c}\int_x {\nabla \Omega_T(x) \cdot 
\nabla \Omega_T(x)} } \nonumber \\
&&\times \prod_{i>j}(z_i - z_j)^{2n+1} \prod_{i=1}^{N}[ \varphi_{\sigma_i}(x_i)
e^{\frac{e}{\hbar c}\{\Omega_T(x_i)-i\Omega_L(x)\} } ]
\end{eqnarray}

\section{The observables}
\label{obscal}

We first evaluate the norm, ${\cal N}[W,\Omega_T]$, in the 
hydrodynamic approximation, {\it i.e.}, in the small amplitude and long 
wavelength limit. It is given by,
\begin{eqnarray}
{\cal N}[W,\Omega_T]&~=~& \prod_{i=1}^{N} \int_{x_i} \sum_{\sigma_i} 
|\psi_N(\{x_i,\sigma_i\})|^2 \nonumber \\
&~=~&const\times 
e^{-\int_x  {\bar{W}_{\sigma}(x)W_{\sigma}(x)} e^{-\frac{2e}{\hbar c} 
\{\Omega_T(x)-\bar{\Omega}_T(x) \}}} e^{-\frac {\kappa}{\hbar c}\int_x {\nabla 
\Omega_T(x) \cdot \nabla \Omega_T(x)} } \nonumber \\ 
&& \times \prod_{i=1}^{N} \int_{x_i}
e^{ \sum_i \ln{\bar{W}_{\sigma}(x_i)W_{\sigma}(x_i)} + 
 2\frac{e}{\hbar c} \sum_i\bar{\Omega}_T(x_i) + (2n+1)\sum_{i,j}\ln|x_i-x_j| } 
\end{eqnarray}
As in the case of the Laughlin wave function, the norm has the form of a
classical partition function of a 2D plasma. Here, there is also an "exernal
potential" which is a function of $W^\dagger(x)W(x)$ and $\Omega_T(x)$. In the 
hydrodynamic limit, we write this partition function as a functional integral
over the density field and evaluate it using the saddle point approximation. 
So we change the variables from $\{x_i\}\rightarrow\tilde{\rho}$, where, 
\begin{equation}
\tilde{\rho}(x)=\sum_{i=1}^{N}\delta(x-x_i)
\end{equation}
then for any function $\cal F$ of $\{x_i\}$, we have
\begin{equation}
\frac{1}{N!}\prod_{i=1}^{N} \int dx_i {\cal F}(\{x_i\})~=~
\int {\cal D}[\tilde{\rho}] J[\tilde{\rho}] {\cal F}[\tilde{\rho}]
\end{equation} 
where the jacobian of the transformation is the entropy factor,
\begin{equation}
J[\tilde{\rho}]= e^{\int_x[\tilde{\rho}(x)- 
\tilde{\rho}(x)\ln{\tilde{\rho}(x)}]}
\end{equation}
Hence the norm can be written as,
\begin{eqnarray}
{\cal N}[W,\Omega_T]&~=~&{\it const} \times
e^{-\int_x  {\bar{W}_{\sigma}(x)W_{\sigma}(x)} e^{-\frac{2e}{\hbar c} 
\{\Omega_T(x)-\bar{\Omega}_T(x) \}}} 
e^{-\frac{\kappa}{\hbar c}\int_x \nabla \Omega_T(x) \cdot\nabla \Omega_T(x)}
\nonumber \\
&&\times \int {\cal D}[\tilde{\rho}] e^{-{\cal S}[\tilde{\rho};W]}
\end{eqnarray} 
where,
\begin{eqnarray}
{\cal S}[\tilde{\rho};W]&~=~&
\int_x [-\tilde{\rho}(x)+\tilde{\rho}(x)\ln{\tilde{\rho}(x)}- 
\tilde{\rho}(x)\ln \{ \bar{W}_{\sigma}(x)W_{\sigma}(x) \}-
2\frac{e}{\hbar c} \tilde{\rho}(x)\bar{\Omega}_T(x)
\nonumber \\
&&- 2\pi(2n+1)\tilde{\rho}(x) \frac{1}{\nabla^2}\tilde{\rho}(x)]   
\end{eqnarray}
We evaluate this functional integral in the saddle point limit. Dropping 
$\tilde{\rho} \ln \tilde{\rho}$ term in comparision with 
$\tilde{\rho}\frac{1}{\nabla^2} \tilde{\rho}$ and substituting the
solution of the saddle point equation,
\begin{equation}
\tilde{\rho}(x) = \bar{\rho} -
\frac {1}{4\pi(2n+1)} \nabla^2 \ln \{ \bar{W}_{\sigma}(x)W_{\sigma}(x)\} 
\end{equation}
we get equation (\ref{normexp}).

We evaluate the values of density and spin density by proceeding 
along similar calculational steps employed in evaluating the norm. 
the density is given by, 
\begin{eqnarray}
\rho(x) &\equiv&  \langle W|{\hat \rho}(x)|W\rangle \nonumber \\
&~=~&
\frac{1}{{\cal N}(W,\Omega_T)}
\prod_{i=1}^{N} \left(\int dx_i \sum_{\sigma_i} \right) 
\bar{\psi}_N(\{x_i,\sigma_i\})\sum_{i=1}^{N} \delta (x -x_i)    
\psi_N(\{x_i,\sigma_i\}) \nonumber \\
&=& \frac{1}{{\cal Z}}
\int {\cal D}[\tilde{\rho}] \tilde{\rho}(x)e^{-{\cal S}[\tilde{\rho};W]}
\nonumber \\
&~\equiv~& \langle \tilde{\rho}(x) \rangle \nonumber \\
&~=~& \frac {-1}{4\pi(2n+1)} \nabla^2 \ln \{ \bar{W}_{\sigma}(x)W_{\sigma}(x)\} + 
\bar{\rho} 
\end{eqnarray}
where $ {\cal Z} \equiv \int {\cal D}[\tilde{\rho}] e^{-{\cal S}[\tilde{\rho};W]}$ 

The spin density is,
\begin{eqnarray}
s^a(x) &~\equiv~&  \langle W|{\hat s}^a(x)|W\rangle \nonumber \\
&~=~& \frac{1}{{\cal N}(W,\Omega_T)}
\prod_{i=1}^{N} \left(\int dx_i \sum_{\sigma_i} \right) 
\bar{\psi}_N(\{x_i,\sigma_i\})\sum_{i=1}^{N} \frac{1}{2}\tau(i)\delta (x -x_i)    
\psi_N(\{x_i,\sigma_i\}) \nonumber \\
&~=~& \frac{1}{2}\frac {\bar{W}_{\sigma}(x)\tau_{\sigma \sigma'}W_{\sigma'}(x)}
{ \bar{W}_{\sigma''}(x)W_{\sigma''}(x) } \langle \tilde{\rho}(x) \rangle
\end{eqnarray}

The current density is,
\begin{eqnarray}
J_j(x) &~\equiv~&  \langle W|{\hat J}_j(x)|W\rangle \nonumber \\
&~=~& \frac{1}{{\cal N}(W,\Omega_T)}
\prod_{i=1}^{N} \left(\int dx_i \sum_{\sigma_i} \right) 
\bar{\psi}_N(\{x_i,\sigma_i\})\sum_{i=1}^{N} \frac{1}{2}\delta (x -x_i)
[-i\hbar\partial_{{x_i}_j} - \frac{e}{c}A_j(x_i)]  \psi_N(\{x_i,\sigma_i\}) 
\nonumber \\
&~+~& h.c 
\nonumber \\
&~=~& \rho(x)[\hbar L_j^3(x) - \frac{e}{c}(A_j(x)-\alpha_j(x))]
\end{eqnarray}
where $i$ is the particle index and $j$ is the coordinate index. $\alpha$ is 
defined through the relation 
$\kappa \nabla \times {\vec \alpha}(x) = {\it e} \rho(x)$
and $L^3_j \equiv \frac{1}{2i}(Z^\dagger \partial_j Z-h.c)$ for 
$j = 1$ and $2$. 

The kinetic energy density is, 
\begin{eqnarray}
{\cal T}(x) &\equiv&  \langle W|{\hat {\cal T}}(x)|W\rangle 
\nonumber \\
&~=~& \frac{1}{{\cal N}(W,\Omega_T)}
\prod_{i=1}^{N} \left(\int dx_i \sum_{\sigma_i} \right) 
\sum_{i=1}^{N}     
|D_i\psi_N(\{x_i,\sigma_i\})|^2 \nonumber \\
&~=~& \hbar \omega_c 
\frac {\partial_z\bar{W}_{\sigma}(x) \partial_{\bar{z}}W_{\sigma}(x)}
{ \bar{W}_{\sigma'}(x)W_{\sigma'}(x) } \langle \tilde{\rho}(x) \rangle
\end{eqnarray}
where $D_i = \partial_{\bar{z}_i} + \frac{1}{2}z_i$ and $z=\frac{x_1+i x_2}{l_c\sqrt{2}}$.

\section{The $\Omega_T$ integral}
\label{omegatint}

In this appendix we evaluate ${\cal G}[W]$ by doing the  $\Omega_T$ integral 
in equation (\ref{gdef}). The saddle-point approximation of the integral gives
\begin{equation}
{\cal G}[W] =  const \times e^{-\int_x \frac{4e}{\hbar c}\tilde{\Omega}_T(x)} 
{\cal N}[W,\tilde{\Omega}_T]
\end{equation}
where $\tilde{\Omega}_T$ is the solution to the saddle-point equation which, 
in long wavelength limit  ($\nabla^2  \ln {\bar{W}W } \ll \ln {\bar{W}W }$)
is,
\begin{equation}
\tilde{\Omega_T}(x) = \bar{\Omega_T}(x) +
\frac{\hbar c} {2e}\ln \{ \bar{W}_{\sigma}(x)W_{\sigma}(x) \}
\end{equation}
When this value for $\tilde{\Omega_T}$ is substituted in the above equation 
we get equation (\ref{gexp}).

\section{The overlaps}
\label{olapcal}
  
The overlap of two gauge invariant coherent states 
$|W_1\rangle$ and $|W_2\rangle$,
obtained by proceeding with steps similar to those involved in evaluating the 
norm, is  
\begin{eqnarray}
\langle W_1|W_2\rangle &=& \frac{1}{\sqrt{{\cal N}(W_1,\Omega_{T1}){\cal N}(W_2,\Omega_{T2})}}
\prod_{i=1}^{N} \left(\int dx_i \sum_{\sigma_i} \right) 
    \bar{\psi}_1(\{x_i,\sigma_i\})\psi_2(\{x_i,\sigma_i\}) \nonumber \\
&=&e^{-\frac{1}{8\pi(2n+1)}\int_x [f_{12}(x)\nabla^2 f_{12}(x)
-\frac{1}{2} f_{11}(x)\nabla^2 f_{11}(x)
 - \frac{1}{2} f_{22}(x)\nabla^2 f_{22}(x)]}
\label{eq:ol1}
\end{eqnarray}
where 
\begin{equation}
f_{ab}(x) = \ln \{ \bar{W}_{a\sigma}(x)W_{b\sigma}(x) \} + 
2\frac{e}{\hbar c} \bar{\Omega}_T(x)
\end{equation}
If we express $W$ in terms of $\rho$ and $Z$ we get the overlap to be 
\begin{eqnarray}
\langle W_1|W_2\rangle
&~=~&e^{-\frac{1}{8\pi(2n+1)}\int_x 
\ln \{ Z_1^\dagger(x)Z_2(x)\}\nabla^2 \ln \{Z_1^\dagger(x)Z_2(x)\}}
\nonumber \\
&&\times e^{\frac{1}{2}\int_x(\rho_1(x)+\rho_2(x))\ln\{Z_1^\dagger(x)Z_2(x)\}}
\nonumber \\
&& e^{ \frac{1}{2}\pi(2n+1)\int_x(\rho_1(x)-\rho_2(x))\frac{1}{\nabla^2}
(\rho_1(x)-\rho_2(x))}
\label{eq:ol2}
\end{eqnarray}
The first term in the exponent of the RHS can be dropped with respect to the 
second one in the long wavelength limit. 
Making use of the relation, $\bar{Z}_{1\sigma}Z_{2\sigma} = 
e^{\frac{{\it i}}{2}\Phi(\vec{n}_1,\vec{n}_2)}
(\frac{1+\vec{n}_1\cdot\vec{n}_2}{2})^{\frac{1}{2}}$ where
$\Phi(\vec{n}_1,\vec{n}_2)$ is the area of the spherical triangle
with vertices at $\vec{n}_1,\vec{n}_2$ and a third point on the unit sphere, 
in the above equation we get the equation (\ref{olapexp}).

\end{appendix}


\begin{references}

\bibitem{tapash}
T. Chakraborty and F. C. Zhang,  
Phys. Rev. {\bf B29}, 7032 (1984). 

\bibitem{klee}
D. H. Lee and C. L. Kane, Phys. Rev. Lett. {\bf 64}, 1313 (1990).

\bibitem{sondhi}
S. L. Sondhi, A. Karlhede, S. A. Kivelson and E. H. Rezayi, 
Phys. Rev. {\bf B47}, 16419 (1993). 

\bibitem{fertig}
H. A. Fertig, L. Brey, R. C\^ot\'e and A. H. MacDonald,  Phys. Rev.
{\bf B50}, 11018 (1994)\,; K. Moon, {\it et. al.}, 
{\it ibid.} {\bf 51}, 5138 (1995).

\bibitem{barret}
S. E. Barrett, G. Dabbagh, L. N. Pfeiffer, K. W. West and R.
Tycko, Phys. Rev. Lett. {\bf 74}, 5112 (1995)\,; {\it ibid.} 
{\bf 72}, 1368 (1994)\,; R. Tycko, S. E. Barrett, G. Dabbagh,
L. N. Pfeiffer and K. W. West, Science {\bf 268}, 1460 (1995).

\bibitem{goldberg}
E. H. Aifer, B. B. Goldberg and D. A. Broido, Phys. Rev. Lett. 
{\bf 76}, 680 (1996).

\bibitem{schrieffer}
A. Schmeller, J. P. Eisenstien, L. N. Pfeiffer and K. W. West, 
Phys. Rev. Lett. {\bf 75}, 4290 (1995).

\bibitem{press}
D.\ K.\ Maud {\it et. al.}, Phys.\ Rev.\ Lett.\ {\bf 77}, 4604 (1996).

\bibitem{shaygan}
S. P. Shukla, M.Shayegan, S. R. Parihar, S. A. Lyon, N. R. Cooper and A. A. Kiselev, 
Phys. Rev. {\bf B61}, 4469 (2000). 

\bibitem{kukushkin}
I. V. Kukushkin, K. v. Klitzing and K. Eberi, 
Phys. Rev. {\bf B55}, 10607 (1997). 

\bibitem{bychkov}
Yu. A. Bychkov, T. Maniv and I. D. Vagner, JETP Lett. {\bf 62}, 
727 (1995).

\bibitem{rray}
Rashmi Ray,
Phys. Rev. {\bf B60}, 14154 (1999). 

\bibitem{girvin}
M. Abolfath, J. J. Palacios, H. A. Fertig, S. M. Girvin and A. H. MacDonald
Phys. Rev. {\bf B56}, 6795 (1997). 

\bibitem{kivelson}
S. C. Zhang, H. Hanson and S. Kivelson, Phys. Rev. Lett. {\bf 62}, 82 (1989).

\bibitem{gmrs}
Ganapathy Murthy and R. Shankar,
Phys. Rev. {\bf B59}, 12260 (1999). 

\bibitem{ezawa}
Z. F. Ezawa,
Physics Letters {\bf A249}, 223 (1998). 

\bibitem{perel}
A. Perelomov, Generalized Coherent States and Their Applications,
Springer-Verlag (1986)

\end{references}
\end{document}